\documentclass[12pt]{article}
\usepackage{graphicx}
\def\nn{\noindent}
\def\Re{{\cal R \mskip-4mu \lower.1ex \hbox{\it e}\,}}
\def\Im{{\cal I \mskip-5mu \lower.1ex \hbox{\it m}\,}}
\def\ie{{\it i.e.}}
\def\eg{{\it e.g.}}
\def\etc{{\it etc}}

\def\sub#1{_{\lower.25ex\hbox{$\scriptstyle#1$}}}
\def\tev{\,{\ifmmode\mathrm {TeV}\else TeV\fi}}
\def\gev{\,{\ifmmode\mathrm {GeV}\else GeV\fi}}
\def\mev{\,{\ifmmode\mathrm {MeV}\else MeV\fi}}
\def\mpl{\ifmmode M_{pl}\else $M_{pl}$\fi}
\def\to{\rightarrow}

\def\subw{_{\rm w}}
\def\mh{\ifmmode m\sbl H \else $m\sbl H$\fi}
\def\mch{\ifmmode m_{H^\pm} \else $m_{H^\pm}$\fi}
\def\mt{\ifmmode m_t\else $m_t$\fi}
\def\mc{\ifmmode m_c\else $m_c$\fi}
\def\mz{\ifmmode M_Z\else $M_Z$\fi}
\def\mw{\ifmmode M_W\else $M_W$\fi}
\def\mws{\ifmmode M_W^2 \else $M_W^2$\fi}
\def\mhs{\ifmmode m_H^2 \else $m_H^2$\fi}   
\def\mzs{\ifmmode M_Z^2 \else $M_Z^2$\fi}
\def\mts{\ifmmode m_t^2 \else $m_t^2$\fi}
\def\mcs{\ifmmode m_c^2 \else $m_c^2$\fi}
\def\mchs{\ifmmode m_{H^\pm}^2 \else $m_{H^\pm}^2$\fi}
\def\ztwo{\ifmmode Z_2\else $Z_2$\fi}
\def\zone{\ifmmode Z_1\else $Z_1$\fi}
\def\mtwo{\ifmmode M_2\else $M_2$\fi}
\def\mone{\ifmmode M_1\else $M_1$\fi}
\def\tb{\ifmmode \tan\beta \else $\tan\beta$\fi}
\def\xw{\ifmmode x\subw\else $x\subw$\fi}
\def\ch{\ifmmode H^\pm \else $H^\pm$\fi}
\def\lum{\ifmmode {\cal L}\else ${\cal L}$\fi}
\def\inpb{\,{\ifmmode {\mathrm {pb}}^{-1}\else ${\mathrm {pb}}^{-1}$\fi}}
\def\infb{\,{\ifmmode {\mathrm {fb}}^{-1}\else ${\mathrm {fb}}^{-1}$\fi}}
\def\epem{\ifmmode e^+e^-\else $e^+e^-$\fi}
\def\ppb{\ifmmode \bar pp\else $\bar pp$\fi}
\def\bsg{\ifmmode B\to X_s\gamma\else $B\to X_s\gamma$\fi}
\def\bsll{\ifmmode B\to X_s\ell^+\ell^-\else $B\to X_s\ell^+\ell^-$\fi}
\def\bstt{\ifmmode B\to X_s\tau^+\tau^-\else $B\to X_s\tau^+\tau^-$\fi}
\def\lamt{\ifmmode \tilde\lambda\else $\tilde\lambda$\fi}
\def\shat{\ifmmode \hat s\else $\hat s$\fi}
\def\that{\ifmmode \hat t\else $\hat t$\fi}
\def\uhat{\ifmmode \hat u\else $\hat u$\fi}

\newskip\zatskip \zatskip=0pt plus0pt minus0pt
\def\matth{\mathsurround=0pt}
\def\lsim{\mathrel{\mathpalette\atversim<}}
\def\gsim{\mathrel{\mathpalette\atversim>}}
\def\atversim#1#2{\lower0.7ex\vbox{\baselineskip\zatskip\lineskip\zatskip
  \lineskiplimit 0pt\ialign{$\matth#1\hfil##\hfil$\crcr#2\crcr\sim\crcr}}}

\renewcommand{\thefootnote}{\fnsymbol{footnote}}

\begin{document} \begin{titlepage} 
\rightline{\vbox{\halign{&#\hfil\cr
&SLAC-PUB-9564\cr
&November 2002\cr}}}
\begin{center}

{\Large\bf Transverse Polarization Signatures of Extra Dimensions 
at Linear Colliders}
\footnote{Work supported by the Department of 
Energy, Contract DE-AC03-76SF00515}
\medskip

\normalsize 
{\bf \large Thomas G. Rizzo}
\vskip .3cm
Stanford Linear Accelerator Center \\
Stanford University \\
Stanford CA 94309, USA\\
\vskip .2cm

\end{center}

\begin{abstract}
If significant longitudinal polarization of both the electrons and positrons  
becomes feasible at a future linear collider(LC), it may be possible to use 
spin rotators to produce transversely polarized beams. Using the transverse 
polarization of both beams, new azimuthal spin asymmetries can be formed which 
will be sensitive probes for new physics beyond the Standard Model. Here we 
demonstrate that these asymmetries are particularly sensitive to the exchange 
of Kaluza-Klein towers of gravitons, or other spin-2 fields, that are 
predicted to exist in higher dimensional theories which address the hierarchy 
problem. These new asymmetries are shown to be able to extend the search reach 
for such new physics by more than a factor of two, provide an additional tool 
for isolating the signatures for spin-2 exchange up to mass scales in excess 
of $10\sqrt s$, and can be used to help differentiate among the proposed 
solutions to the hierarchy problem below the production threshold for new 
particles. 
\end{abstract} 



\renewcommand{\thefootnote}{\arabic{footnote}} \end{titlepage}


\section{Introduction}

New Physics beyond the Standard Model (SM) is expected to lie at or near the 
TeV scale on rather general grounds. Once this scale is probed by future 
colliders, such at the LHC and the Linear Collider (LC), this new physics 
should begin to show itself. What is uncertain is the form this 
manifestation will take. The most straightforward scenario to visualize and 
analyze experimentally would be the production of new particles such as SUSY or 
Kaluza-Klein resonances. A second possibility is that new processes 
which are not allowed 
within the SM framework will begin to be observed; it may be very difficult 
in this case to access the underlying theory. Lastly, one can imagine  
that the data begins show small deviations from the SM predictions 
for various observables, \eg, cross sections and asymmetries, which 
grow with increasing energy. This last possibility signals the existence of 
new physics beyond the kinematic reach of the collider which is manifesting 
itself in the form of higher dimensional operators, \ie, generalized 
contact interactions. These operators can 
arise from the exchanges of new particles, too massive to be directly 
produced, with different spins and in various channels depending upon the 
particular model. From the literature it is easy to construct a rather long 
list of potential new physics scenarios of this type; clearly this list does 
not exhaust all of the possibilities: a $Z'$ from an extended 
gauge model{\cite {e6,zp}}, scalar or vector leptoquarks{\cite {e6,lq}, 
$R$-parity violating sneutrino($\tilde \nu$) exchange{\cite {rp}}, scalar or 
vector bileptons{\cite {bl}}, graviton Kaluza-Klein(KK) 
towers{\cite {ed,dhr}} in extra dimensional models{\cite {add,rs}}, 
gauge boson KK towers{\cite {ed2,dhr}}, and even string 
excitations{\cite {se}}. 

If such deviations are observed it will be necessary to 
have techniques 
available to differentiate the multiple possibilities experimentally 
and point us in the direction of the correct scenario. One possible path to 
take if this situation is realized is to compare the observed shifts with the 
predictions of all of the currently available models{\cite {gaby}}. An 
alternative to this approach is to develope specific tools to rapidly 
identify certain classes of models which lead to uniquely distinct signatures. 
In this paper we 
examine one such tool which becomes available at the LC provided both 
the $e^-$ and $e^+$ beams are initially 
longitudinally polarized and spin rotators are 
used to convert these to transversely polarized beams. As we will see below, 
transverse polarization (TP){\cite {tp}} allows for new asymmetries to be 
constructed which are associated with the azimuthal angle formed by the 
directions of the $e^\pm$ polarization and the plane of the momenta of the 
outgoing fermions  
in the $e^+e^-\to f\bar f$ process. Historically, the possible 
use of TP as a tool for new physics searches and analyses 
has not gotten the attention it deserves 
in the literature{\cite {tp}}. Here, in an effort to partially remedy this 
situation, we are interested in using the associated TP 
asymmetries to uniquely probe for the $s$-channel exchange of spin-2 fields 
in $e^+e^-$ collisions 
which we normally associate with the Kaluza-Klein graviton towers of the 
Arkani-Hamed, Dimopoulos and Dvali(ADD){\cite {add}} or 
Randall-Sundrum(RS){\cite {rs}} scenarios. 

The organization of this paper is as follows. After our introduction, we 
outline the influence of transverse polarization on the process 
$e^+e^- \to f\bar f$ in the presence of graviton exchange in both the ADD and 
RS scenarios. In particular we examine how the new asymmetries associated with 
transverse polarization can be used to probe for spin-2 
graviton-like exchanges. 
In section 3 we analyze these asymmetries in detail and their applications for 
spin-2 exchange identification at a future LC. We demonstrate the usefulness 
of TP in distinguishing $s$-channel spin-2 exchange from other possible new 
physics scenarios. We will see that TP allows one to expand the 
sensitivity range for the 
cutoff scale over which graviton KK exchange can be observed by up to a factor 
of two. We also show that TP may allow us to differentiate the ADD 
and RS scenarios below KK production threshold. Our conclusions can be 
found in Section 4.

\section{Transverse Polarization Asymmetries}

For our analysis we will follow a slightly modified version of the notations 
and conventions employed by  Hikasa{\cite {tp}}. Consider the process 
$e^+e^- \to f\bar f$ with the both electron and positron beams polarized. 
We will denote the linear and transverse components of the $e^-(e^+)$ 
polarizations by $P_{L,T}(P_{L,T}')$ and for simplicity assume that the two 
transverse polarization vectors are parallel up to a sign. In this case, the 
spin-averaged matrix element for this process can be written as
\begin{eqnarray}
|\bar {\cal M}|^2&=&{1\over {4}}(1-P_L P_L')(|T_+|^2+|T_-|^2)+(P_L-P_L')
(|T_+|^2-|T_-|^2)\, \nonumber \\
&+&(2P_T P_T')[\cos 2\phi ~Re(T_+T_-^*)-\sin 2\phi ~Im(T_+T_-^*)]\,,
\end{eqnarray}
where $\phi$ is the azimuthal angle defined on an event-by-event basis 
described above. (Note that $\phi$ is defined in such a way so 
that the $2\to 2$ process 
always takes place in the $\phi=0$ plane.) It is interesting to note that the 
$\phi-$dependent pieces of $|\bar {\cal M}|^2$ 
are particularly sensitive to the relative phases between the two sets of 
amplitudes. We also observe from this expression the important fact that 
the $\phi$-dependent pieces are {\it only} accessible if both 
beams are simultaneously 
transversely polarized. Thus to have azimuthal asymmetries at a LC 
we must begin with both beams longitudinally polarized and employ spin 
rotators. 

Let us first consider the simple case 
with massless fermions in the final state assuming no $s-$channel scalar 
exchanges are present. Let us define the quantities  
\begin{eqnarray}
f_{LL} &=& Q_e Q_f+g_Z (v_e-a_e)(v_f-a_f)P \, \nonumber \\
f_{RR} &=& Q_e Q_f+g_Z (v_e+a_e)(v_f+a_f)P \, \nonumber \\
f_{LR} &=& Q_e Q_f+g_Z (v_e-a_e)(v_f+a_f)P \, \nonumber \\
f_{RL} &=& Q_e Q_f+g_Z (v_e+a_e)(v_f-a_f)P \,,
\end{eqnarray}
where $v_e,a_e(v_f,a_f)$ are the vector and axial vector couplings of the 
initial electron(final fermion) 
to the $Z$ and $Q_{e,f}$ are their corresponding electric 
charges, with 
\begin{equation}
g_Z={G_F M_Z^2\over {2\sqrt 2 \pi \alpha}}\,,
\end{equation}
and 
\begin{equation}
P={s\over {s-M_Z^2+iM_Z\Gamma_Z}}\,. 
\end{equation}
Without scalar exchange but allowing for the possibility of 
spin-2 the relevant helicity amplitudes for this process are given by 
\begin{eqnarray}
T_{+-}^{+-}=f_{LL}(1+z)-f_g(z+2z^2-1)\, \nonumber\\
T_{+-}^{-+}=f_{LR}(1-z)-f_g(z-2z^2+1)\, \nonumber\\
T_{-+}^{+-}=f_{RL}(1-z)-f_g(z-2z^2+1)\, \nonumber\\
T_{-+}^{-+}=f_{RR}(1+z)-f_g(z+2z^2-1)\,. 
\end{eqnarray}
where $z=\cos \theta$. 
Note that the spin-2 exchange merely augments the amplitudes which are 
already present in the SM(though with different $\cos \theta$ 
dependencies), \ie, no 
new helicity amplitudes are generated by spin-2 over those due to spin-1. In 
contrast to this, scalar exchange would yield additional 
amplitudes of the form $T_{++}^{++}$ \etc. not present in the SM 
and would thus be easily isolated using the more conventional asymmetries 
associated with two beam longitudinal polarization{\cite {rp}}. Clearly 
isolating spin-2 exchange will be 
somewhat more difficult since no new amplitudes appear. 
$f_g$ is a model-dependent quantity; in the usual ADD model, employing 
the convention of Hewett{\cite {ed}}, one finds 
\begin{equation}
f_g={\lambda s^2\over {4\pi \alpha M_H^4}}\,. 
\end{equation}
where $M_H$ represents the cutoff scale in the KK graviton tower sum 
and $\lambda=\pm 1$. 
In the RS model the corresponding expression can be obtained through the 
replacement
\begin{equation}
{\lambda \over {M_H^4}} \to {-1\over {8\Lambda_\pi^2}}\sum_n 
{1\over {s-m_n^2+im_n\Gamma_n}}
\,. 
\end{equation}
where $\Lambda_\pi$ is of order a few TeV and $m_n(\Gamma_n)$ are the 
masses(widths) of the TeV 
scale graviton KK excitations. In what follows we will always assume that we 
are below the threshold for the production of these resonances otherwise the 
spin-2 nature of the new exchange would be easily identified through an 
examination of the resonances themselves. We will also assume that we are 
sufficiently distant from these resonances in energy that there widths can 
be neglected in cross section calculations. 

In the case of massive final state fermions, such as tops, 
the helicity amplitudes given 
above are slightly altered and new amplitudes $T_{+-}^{\pm \pm}$ and 
$T_{-+}^{\pm \pm}$ are also present. The exact forms of these expressions 
in this case are not very enlightening so we will not present them here 
but they will be included 
in the analysis in the case of top quark pair production. 

What is the form or the angular distribution, $d\sigma/dz d\phi$, in the SM? 
(Here we define $z=\cos \theta$ as above.) In particular, we are interested 
in the $z$-dependence of the terms associated with $\cos 2\phi$ and 
$\sin 2\phi$ in the expression above when no new physics is present. We note 
that the a small `imaginary' term will be present even in the SM due to the 
finite width of $Z$. For $\sqrt s \geq 500$ GeV this term can be safely 
neglected for most of our analyses here but we include it for completeness.  
As shown in, \eg, the work of Hikasa, both of these $\phi$-dependent terms are 
always proportional to $1-z^2$ in the SM and will remain so even if new gauge 
boson exchanges are present. However, due to the more complex $z$-dependence 
of the spin-2 contributions to the helicity amplitudes we expect 
significant modifications 
of the SM result when gravitons are exchanged. In fact, interference between 
SM and spin-2 exchange amplitudes are found to produce both even and odd$-z$ 
terms with the latter  
proportional to $\sim z(1-z^2)$ whereas the smaller pure gravity terms are 
instead found to be even in $z$ and proportional to $z^2-(2z^2-1)^2$. 
The general difference in the $z$-dependence of the of the $\phi$ 
sensitive terms and, in particular, the existence of the odd-$z$ 
contributions is clearly a signal for spin-2 exchange. We note in passing 
that in the case of scalar exchange no odd-$z$ terms will be generated since 
the spin-0 and spin-1 exchanges do not interfere. 

Let us assume, as mentioned above, that we are in an energy regime where the 
effects of the finite width of the $Z$ can be neglected. For the moment, 
this would seem to imply that the term proportional to $\sin 2\phi$ can be 
neglected in the case of KK graviton exchange. Let us proceed making this 
assumption but remembering to return to 
this important point below. We will later see that the terms that we now 
neglecting will have no influence on this part of our analysis. In order 
to attempt to 
isolate the spin-2 exchange contributions we first can form a differential 
azimuthal asymmetry distribution which we symbolically define by 
\begin{equation}
{1\over {N}} {dA\over {dz}}=\Bigg[{{\int_+ {d\sigma \over {dz d\phi}}-\int_- 
{d\sigma \over {dz d\phi}}}\over {\int d\sigma}}\Bigg]\,,
\end{equation}
where $\int_{\pm}$ are integrations over regions where $\cos 2\phi$ takes on 
$\pm$ values; integration over the full ranges of $z$ and $\phi$ occurs 
in the denominator. It is important to note that we expect this differential 
asymmetry to take on rather small numerical 
values since it is normalized to the total cross section and {\it not} to the 
differential cross section at the same value of $z$ as is usually done. 
As we will see below, this particular normalization is most useful in 
isolating the most important aspects of TP physiucs. To get a feeling 
for this asymmetry, we show its behaviour for both the SM and in the ADD 
scenario in Fig.1 at a 500 GeV LC for the final states $f=\mu $ or $\tau,c$ 
and $b$. Note that from here on we will combine results for the $f=\mu$ and 
$\tau$ final states to get added statistics. In this figure we have 
for concreteness assumed that the spin rotators are nearly 
$100\%$ efficient{\cite {gudi}} so that $P_T=0.8$ and $P_T'=0.6$. Note that 
the spin-2 effects are large and in particular the fact that the azimuthal 
asymmetry distribution is no longer symmetric under $z\to -z$ as we might 
expect from the discussion above.

\vspace*{-0.5cm}
\nn
\begin{figure}[htbp]
\centerline{
\includegraphics[width=7cm,angle=90]{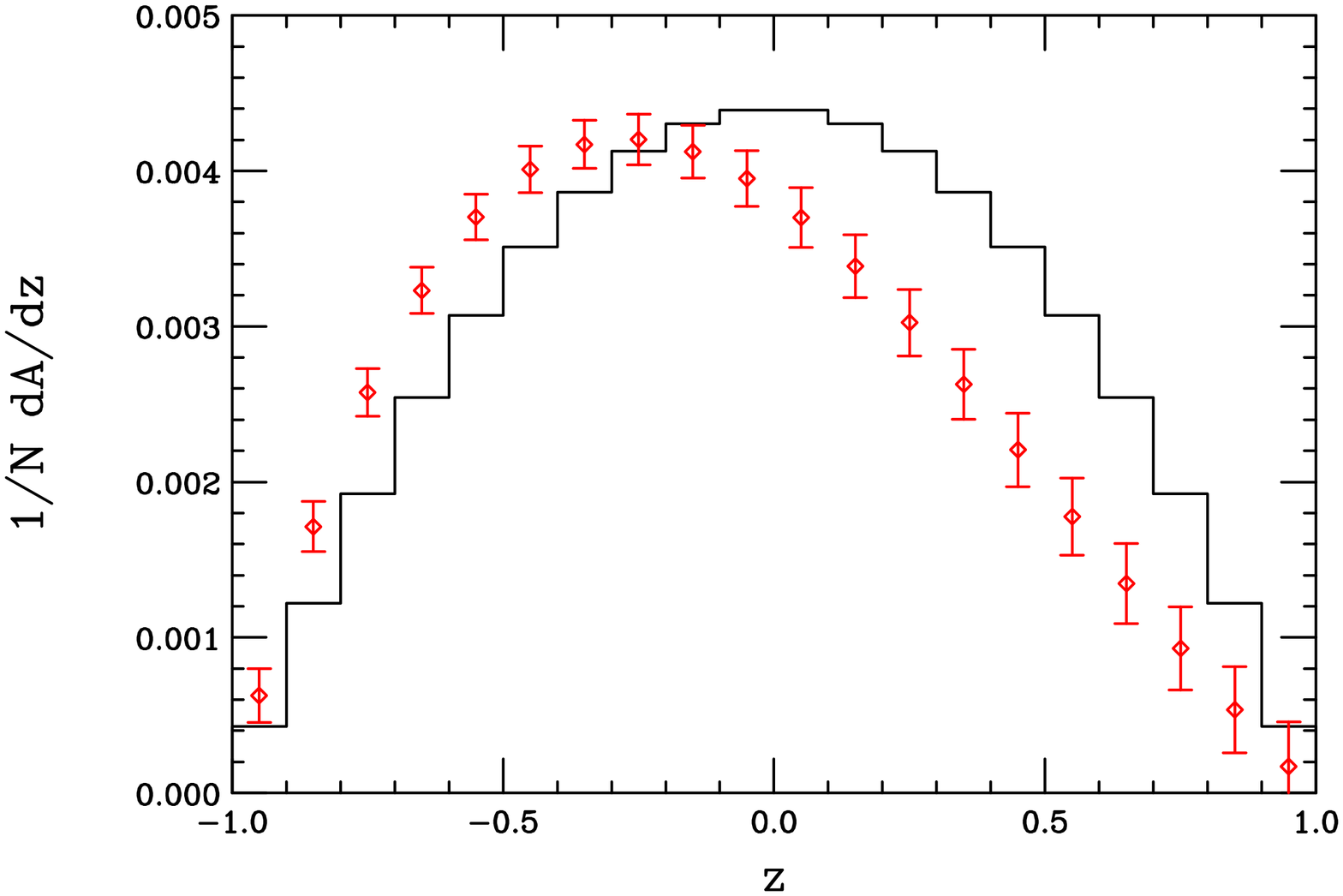}}
\vspace*{0.35cm}
\centerline{
\includegraphics[width=7cm,angle=90]{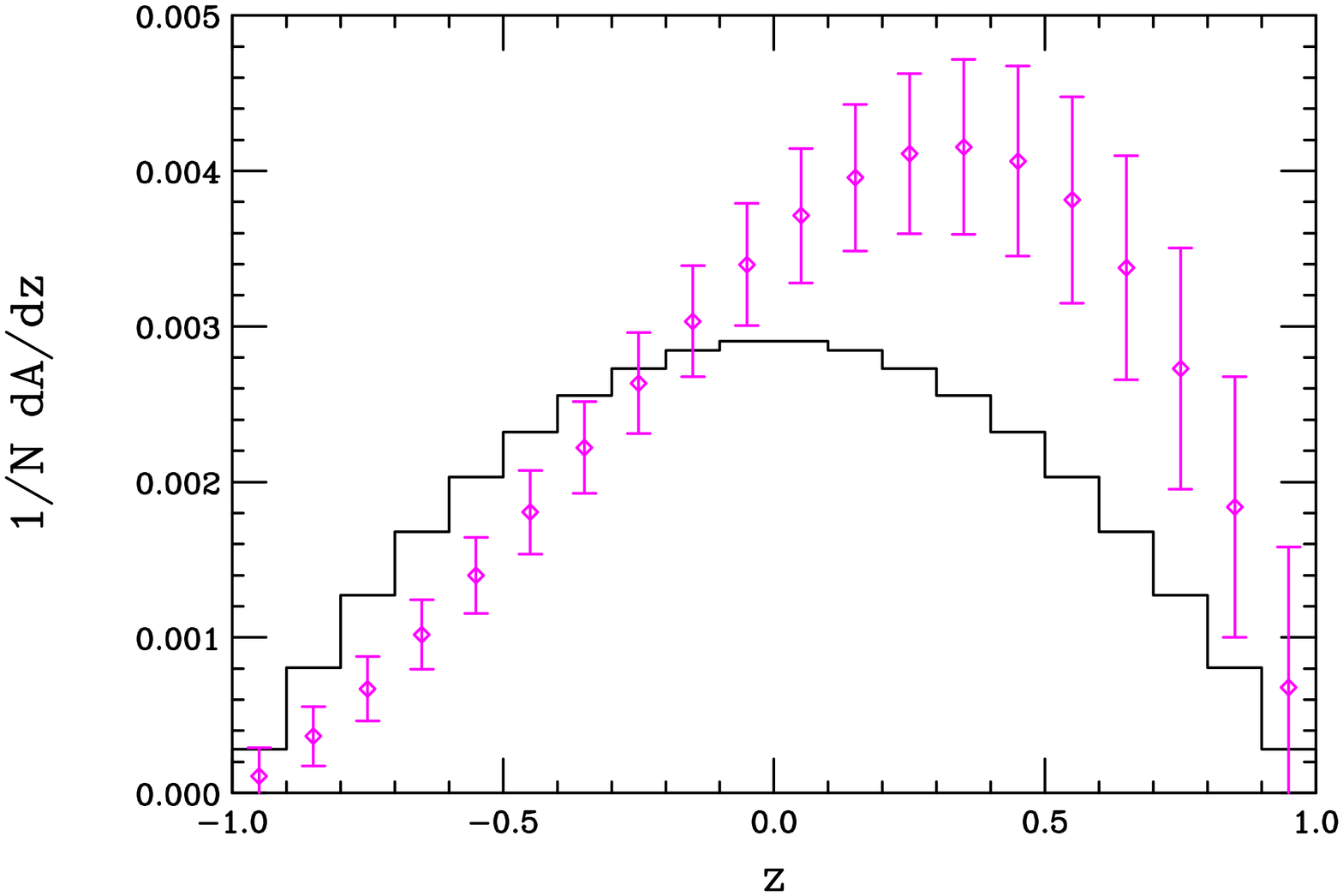}}
\vspace*{0.35cm}
\centerline{
\includegraphics[width=7cm,angle=90]{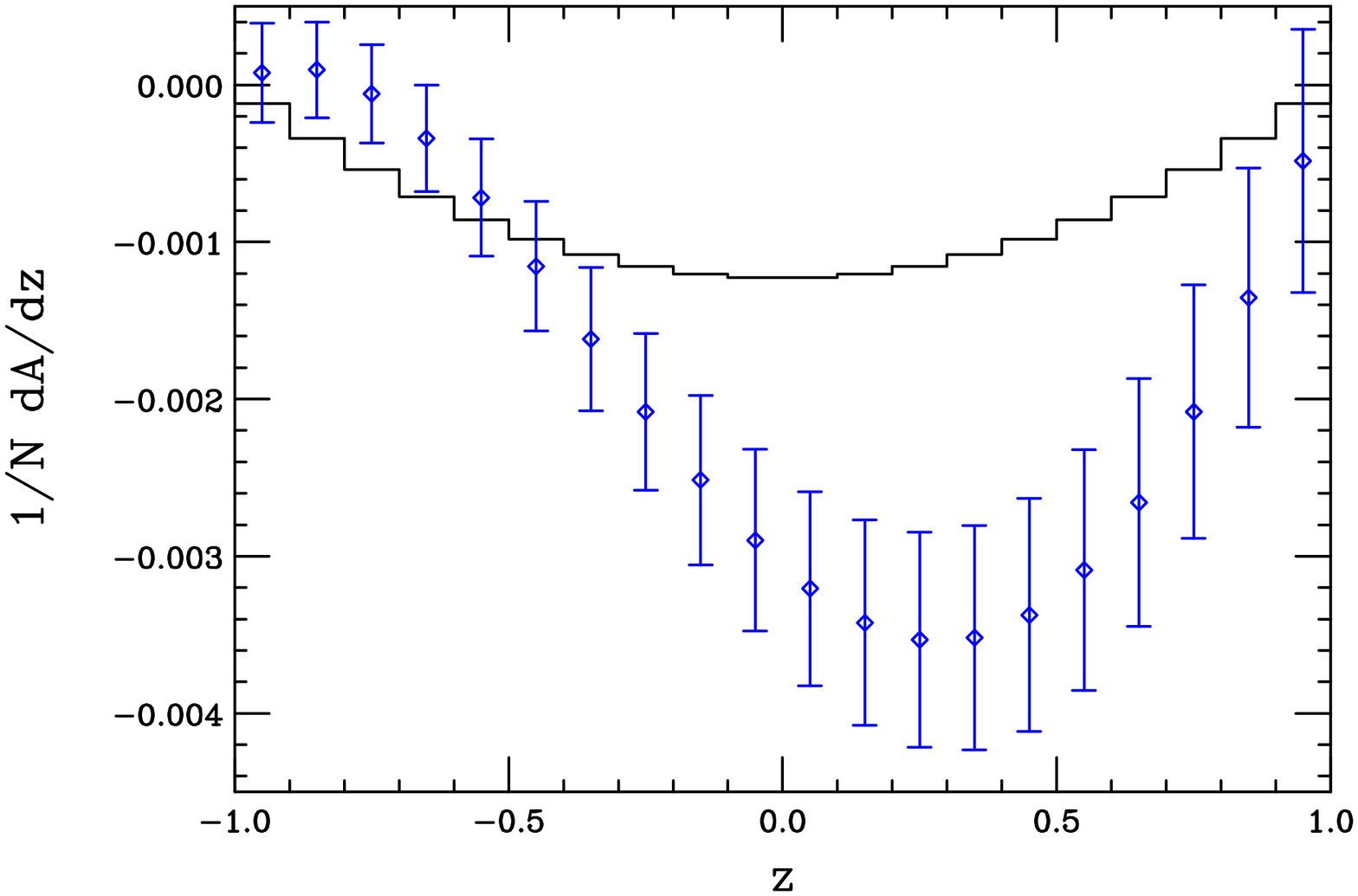}}
\vspace*{0.25cm}
\caption{Differential azimuthal asymmetry distribution for $e^+e^-\to f\bar f$ 
at a 500 GeV LC assuming a luminosity of $500~fb^{-1}$. The histograms are 
the SM predictions while the data points assume the ADD model with $M_H=1.5$ 
TeV. In the top panel $f=\mu$ and $\tau$ are combined, while in the 
middle(lower) panel, $f=c(b)$. $P_T=0.8$ and $P_T'=0.6$ are assumed.}
\end{figure}

There are two ways to naively access the odd-$z$ terms. First, one can take the 
differential azimuthal 
asymmetry defined above, separately integrate it over positive 
and negative values of $z$, then 
take the difference, \ie, form a forward-backward asymmetry using 
$N^{-1} dA/dz$:
\begin{equation}
A_{FB}={1\over {N}} \Bigg[\int_{z\geq 0}~dz~{dA\over {dz}}-\int_{z\leq 0}~dz~
{dA\over {dz}}\Bigg]\,.
\end{equation}
It is important to be reminded that in the SM and in any new physics 
scenario with $s$-channel $Z'$ exchanges one has $A_{FB}=0$. This 
is also true in the 
usual four-fermion contact interaction scenario{\cite {mp}} which involves 
only vector and axial-vector couplings. 
Due to the nature of spin-0 
exchange it is clear that $A_{FB}$ would remain zero in this case as well. 
A second possibility is to take {\it odd} moments of the asymmetry with respect 
to, \eg, the Legendre polynomials $P_n(z)${\cite {tgr}}:
\begin{equation}
<P_n>={1\over {N}} \Bigg[\int~dz~P_n(z){dA\over {dz}}\Bigg]\,. 
\end{equation}
Note that only $<P_{1,3}>$ will be non-zero in this case since no factors of 
$z^5$ appear in the cross section. As in the case of $A_{FB}$, these moments 
are zero in both the SM and $Z'$ models. In the case of graviton exchange, 
not only are the moments $<P_{1,3}>$ non-zero, they are also not independent 
of each other. A short analysis finds that in the case of spin-2 exchange 
the ratio of moments is fixed: 
$<P_{3}>/<P_{1}>= -3/7$, uniquely. It is thus rather obvious that 
the existence of odd-$z$ terms 
is a signal for graviton, or more generally, spin-2 exchange.

\section{Analysis}

It is clear that non-zero values of either $A_{FB}$ or $<P_{1,3}>$ provide a 
clean signature for spin-2 exchange in the $e^+e^- \to f\bar f$ process. 
Their appearance at the level of $5\sigma$ can thus be claimed as, not just 
a discovery of new physics, but spin-2 exchange in particular. 
To be specific in what 
follows let us concentrate on the ADD model; (almost) all limits obtained 
there can be immediately translated to the case of the RS scenario. From Fig.1 
it is apparent that modest values of $M_H$ cause quite 
sizeable distortions in the 
$N^{-1} dA/dz$ distribution. However, as we will see 
this sensitivity is somewhat diluted 
if we are only asking whether or not, \eg  ~$A_{FB}$ is non-zero. After all 
the asymmetry distribution may be quite different than what the SM predicts in 
both magnitude and shape and yet $A_{FB}$ will remain zero. Such a possibility 
will occur in the case of , \eg, spin-0 exchange. To determine 
the $5\sigma$ {\it identification} 
reach we will assume that the individual polarizations 
are known rather well, $\delta P/P=0.003$, that the efficiencies of 
identifying the final state fermions is rather high: $100\%$ for $f=\mu,\tau$, 
$60\%$ for $f=c,t$, and $80\%$ for $f=b$ with no associated systematic 
uncertainties and include the effects of initial 
state radiation. The $5\sigma$ identification reaches, making these 
assumptions, are shown in Figs. 2 and 3 for different values of $\sqrt s$ as 
functions of the integrated luminosity. In obtaining these results 
we have combined all of the various final states above into a single fit. 
In all cases a small angle cut of 
100 mrad around the beam pipe has been employed; our results are not 
particularly sensitive to this value. 

\begin{figure}[htbp]
\centerline{
\includegraphics[width=9cm,angle=90]{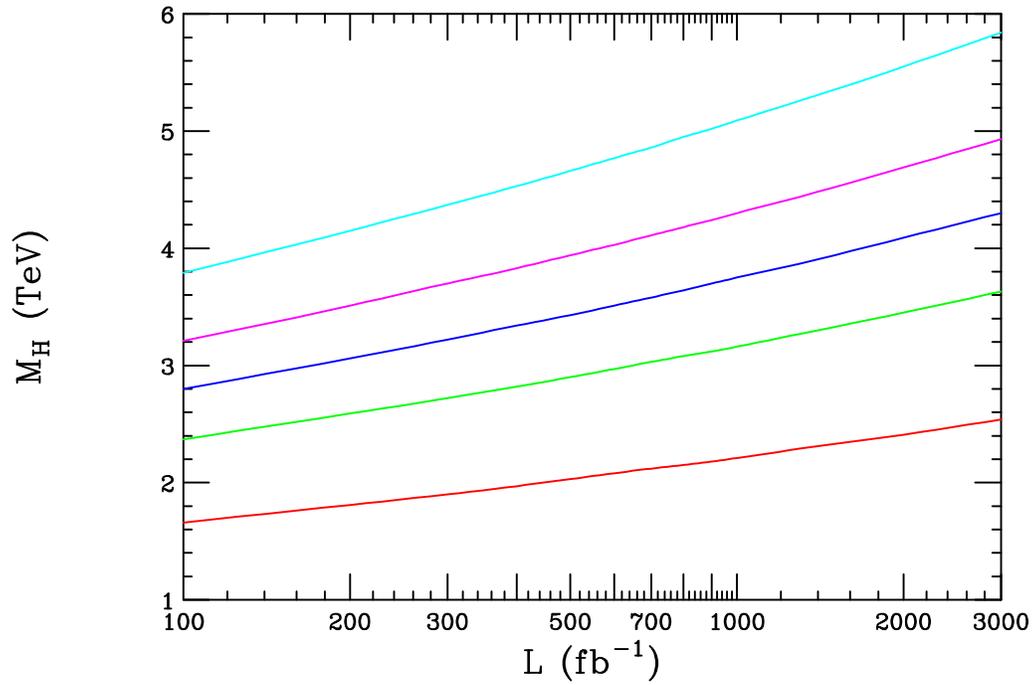}}
\vspace*{0.1cm}
\caption{$5\sigma$ identification 
reach in $M_H$ using $A_{FB}$ as a function of the 
integrated luminosity from the process $e^+e^-\to f\bar f$, with $f$ summed 
over $\mu,\tau,b,c$ and $t$. Here $P_T=0.8$ and $P_T'=0.6$ are assumed. 
From bottom to top 
the curves are for $\sqrt s=0.5, 0.8,1, 1.2$ and $1.5$ TeV, 
respectively.}
\end{figure}

These reaches are, as one might expect, somewhat sensitive to the fact that 
we have assumed values of $P_T=0.8$ and $P_T'=0.6$. If the 
efficiency of the spin rotators is somewhat less than $\sim 100\%$ or if such 
high initial longitudinal polarizations are not achieved the $5\sigma$ 
identification reach will clearly degrade. This reduction in reach will not be 
very serious unless the product $P_TP_T'$ is very greatly reduced or unless 
other errors dominate the experimental uncertainties in the measurements. 

From these two figures some immediate conclusions can be drawn. First, it is 
clear that the identification reach obtained from $A_{FB}$ is somewhat 
superior to that obtained from the measurements of $<P_{1,3}>$. Secondly, it 
is clear that the identification reach in either case alone,  
$M_H \sim (3.5-4)\sqrt s$, is not as good as what can be obtained employing 
longitudinal polarization{\cite {tgr}}. Thirdly, 
it is unclear that combining the two sets 
of observables would be very useful since $A_{FB}$ is correlated with the 
values of $<P_{1,3}>$. In order to obtain better reaches 
we must try something more aggressive.

\begin{figure}[htbp]
\centerline{
\includegraphics[width=9cm,angle=90]{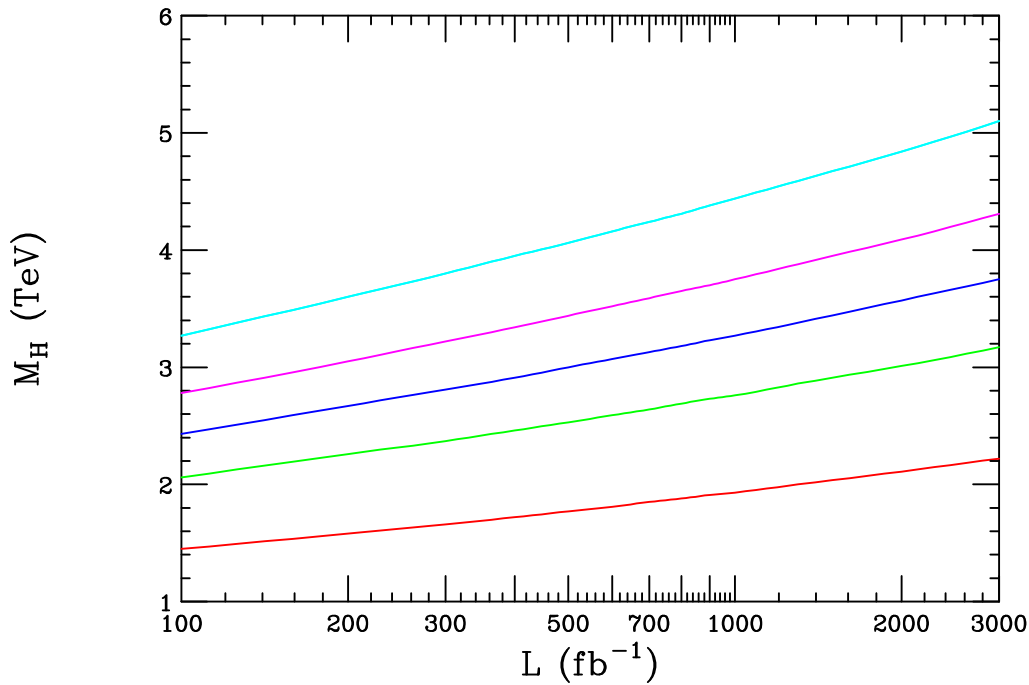}}
\vspace*{0.1cm}
\caption{Same as the previous figure but now using the moments $<P_{1,3}>$.}
\end{figure}

We noted above that in the SM, in all $Z'$ models and in the case of 
conventional four-fermion contact interactions the azimuthal asymmetry always 
takes the form $N^{-1} dA/dz \sim (1-z^2)$. Clearly these 
specific forms of new physics will only modify the normalization of 
the azimuthal asymmetry distribution since its shape is 
left unaltered. We can thus ask up to what value of the cutoff scale, $M_H$, 
can we differentiate the effects of gravity--a change in the {\it shape} of 
these distributions--from a simple overall change in the normalization of 
distributions for the 
various final states. This allows us to set a limit on the value of $M_H$ 
below which graviton exchange can be distinguished from $Z'$ exchange or 
four-fermion contact interactions. To do this we fix $M_H$ and try to fit 
the $N^{-1} dA/dz$ distributions for $\mu,\tau$, $c$ and $b$ final states 
assuming a SM shape but allowing the normalization to float independently 
for each final state. If the CL of the 
fit is very poor we raise $M_H$ until we achieve a CL equivalent to $5\sigma$, 
\ie, $5.7\times 10^{-5}$. For luminosities above $100-200~fb^{-1}$ the errors 
are completely dominated by systematics and we find the results shown in 
Table 1. (Changing the luminosities in our range of interest 
${1\over {2}}-2~fb^{-1}$ has little quantitative effect and 
only modifies the second decimal place in these results.) Here we see that 
for $M_H \lsim (10-11) \sqrt s$ the effects of spin-2 
graviton exchange can be distinguished from a $Z'$ or any form of the 
four-fermion contact 
interactions. This identification reach is numerically similar to the $95\%$ CL 
{\it discovery} reach for graviton exchange obtained 
using only singly longitudinally polarized beams{\cite {tgr,tesla,joa}} for 
the same process. Since 
these results are so dominated by the systematics it is important that a more 
detailed study of this type be performed using a realistic detector 
simulation since the likely size of the true systematic errors will be 
somewhat larger than those assumed in this analysis. However, the reach here 
is so large it is clear that this is an avenue worth persuing. 

\begin{table}[ht]
\begin{center}
\begin{tabular}{|c|c|}
\hline
$E_{CM}$ (GeV)& Reach (TeV) \\ \hline
\hline
500  & 5.4  \\ \hline
800  & 8.8  \\ \hline
1000  & 11.1  \\ \hline
1200  & 13.3  \\ \hline
1500  & 16.7 \\ \hline
\hline
\end{tabular}
\caption{Identification reach for $M_H$ in the ADD model 
assuming the distribution 
$N^{-1} dA/dz \sim 1-z^2$ and 
varying the individual normalizations for the final states $f=\mu,\tau$, 
$f=b$ and $f=c$ for LC of different center of mass energies.}
\end{center}
\end{table}

Given these results we can go a step further. If graviton and $Z'$ 
exchanges can be distinguished up to $M_H\lsim (10-11) \sqrt s$ using TP, 
what is the corresponding $95\%$ CL search reach for graviton 
exchange obtainable with TP?  For this type of analysis we assume that 
the $N^{-1} dA/dz$ distributions for each final state fermion 
are given by their SM values and ask at 
what value of $M_H$ the corresponding ones with graviton exchange become 
indistinguishable from these. Again we find that above very modest integrated 
luminosities the errors are completely dominated by systematics; we thus 
expect our results to again be on the high side of what would be obtained 
in a more detailed detector 
study. These results are shown in Table 2 where we see that the values are 
in the range  
$M_H \gsim 20\sqrt s$. These are such enormous numbers that even a degradation 
by $30-40\%$ would lead to the highest search reaches for KK graviton 
exchange found so far in the literature{\cite {joa}}.

\begin{table}[ht]
\begin{center}
\begin{tabular}{|c|c|}
\hline
$E_{CM}$ (GeV)& Reach (TeV) \\ \hline
\hline
500  & 10.2  \\ \hline
800  & 17.0 \\ \hline
1000  & 21.5  \\ \hline
1200  & 26.0  \\ \hline
1500  & 32.7 \\ \hline
\hline
\end{tabular}
\caption{$95\%$ CL search reach for $M_H$ as described in the text.}
\end{center}
\end{table}

Given the great sensitivity of transverse polarization to KK graviton/spin-2 
exchange it would be natural to ask if TP can be used to distinguish the 
ADD from the RS model scenarios below KK production threshold. At first, 
there would seem to be no difference between 
the predictions of these two models for the situation under discussion. 
In the RS model, if we are away from the $Z$ and 
graviton KK poles the imaginary part of amplitude which enters the term 
proportional to $\sin 2\phi$ becomes vanishingly small. 
However, as was recently pointed out by Datta, Gabrielli and 
Mele{\cite {datta}}, the exchange of an essentially continuous spectrum 
of ADD gravitons leads to a finite, cutoff-independent 
{\it imaginary} part of the amplitude.  This forgotten piece grows very rapidly 
with increasing $\sqrt s$ and depends quite sensitively upon the number of 
extra dimensions. Since this term is finite it directly probes the effective 
fundamental Planck scale of the extra-dimensional theory. Using the notation 
employed above one now finds that $f_g$ has an grown imaginary part: 
\begin{equation}
f_g = {\lambda s^2\over {4\pi \alpha M_H^4}}\Bigg [1-i{\pi M_H^4 (\sqrt s)^
{\delta-2}S_{\delta-1}\over {16 M_D^{\delta+2}}}\Bigg]\,, 
\end{equation}
where $\delta$ is the number of extra dimensions, $M_D$ is the $\delta$ 
dimensional fundamental scale and $S_{\delta-1}$ is the area of the 
$\delta$ sphere. We again note that the magnitude of this new imaginary part, 
unlike the real part as parameterized 
in the Hewett scheme, depends quite strongly on the number of extra dimensions. 

\begin{figure}[htbp]
\centerline{
\includegraphics[width=9cm,angle=90]{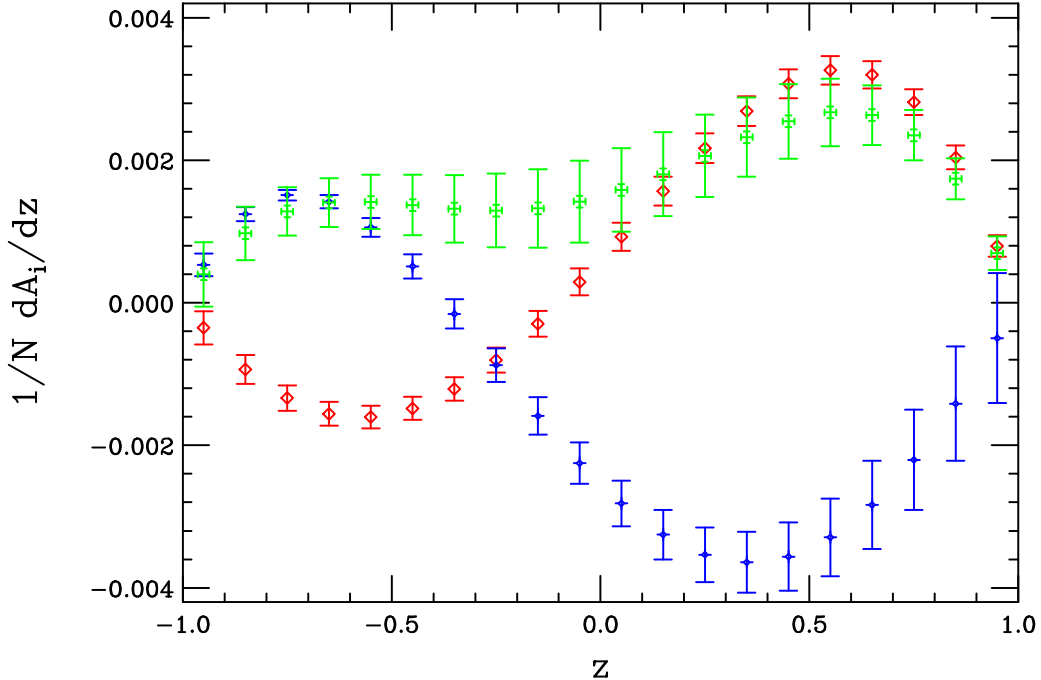}}
\vspace*{0.1cm}
\caption{The $N^{-1} dA_i/dz$ distributions at a 500 GeV collider assuming 
$M_H=M_D=1.5$ TeV and $\delta=3$ with an integrated luminosity of 
500 $fb^{-1}$. The plotted points from top to bottom in the center of the plot 
correspond to $f=b, \mu$ plus $\tau$ and $c$, respectively.}
\end{figure}

To proceed we can form a new asymmetry in analogy to the above: 
\begin{equation}
{1\over {N}} {dA_i\over {dz}}=\Bigg[{{\int_+ {d\sigma \over {dz d\phi}}-\int_- 
{d\sigma \over {dz d\phi}}}\over {\int d\sigma}}\Bigg]\,,
\end{equation}
where now the $\int_{\pm}$ are integrations over regions where $\sin 2\phi$ 
takes on $\pm$ values and we integrate over all $z$ and $\phi$ in the 
denominator 
as before. We note that when we perform the integrations in this manner 
all terms proportional to $\cos 2\phi$ are found to cancel implying that
there is no cross contamination from this other asymmetry source. (This also 
implies that all of our analyses above will go through even if a term 
proportional to $\sin 2\phi$ is present.) Of course this new distribution 
is identically zero in both the SM as well as the RS 
model away from the $Z$ and RS KK graviton poles. Thus, observing {\it any} 
non-zero value for this quantity 
is a signal for the ADD model. This is  particularly true after 
the spin-2 nature of the exchange has already been established. Fig 4 shows 
how these new asymmetry distributions may appear at a 500 GeV LC assuming 
as before that $P_T=0.8$ and $P_T'=0.6$ and taking $\delta=3$ for purposes of 
demonstration. (We do not consider the case $\delta=2$ here as the bounds on 
the scale $M_D$ in this case are in excess of $\sim 100$ TeV{\cite {joa}}). 
For simplicity we have assumed $M_H=M_D$ in this 
figure and will continue to do so in our discussion below; we expect these 
two mass scales to be reasonably comparable, though if for some reason 
$M_H<<M_D$ this would lead to a serious modification in the sensitivity to this 
observable. 

\vspace*{-0.5cm}
\nn
\begin{figure}[htbp]
\centerline{
\includegraphics[width=8cm,angle=90]{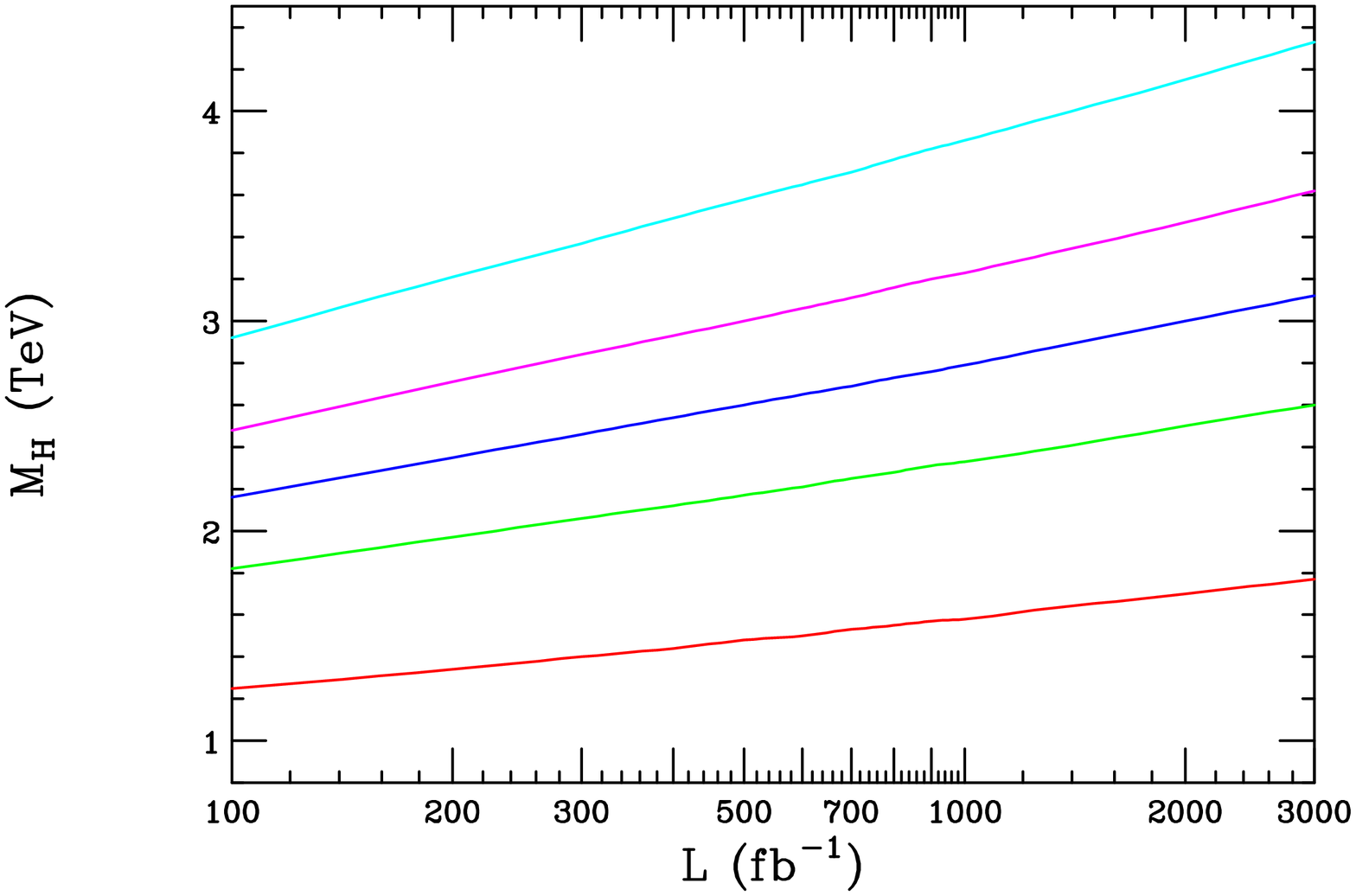}}
\vspace*{0.35cm}
\centerline{
\includegraphics[width=8cm,angle=90]{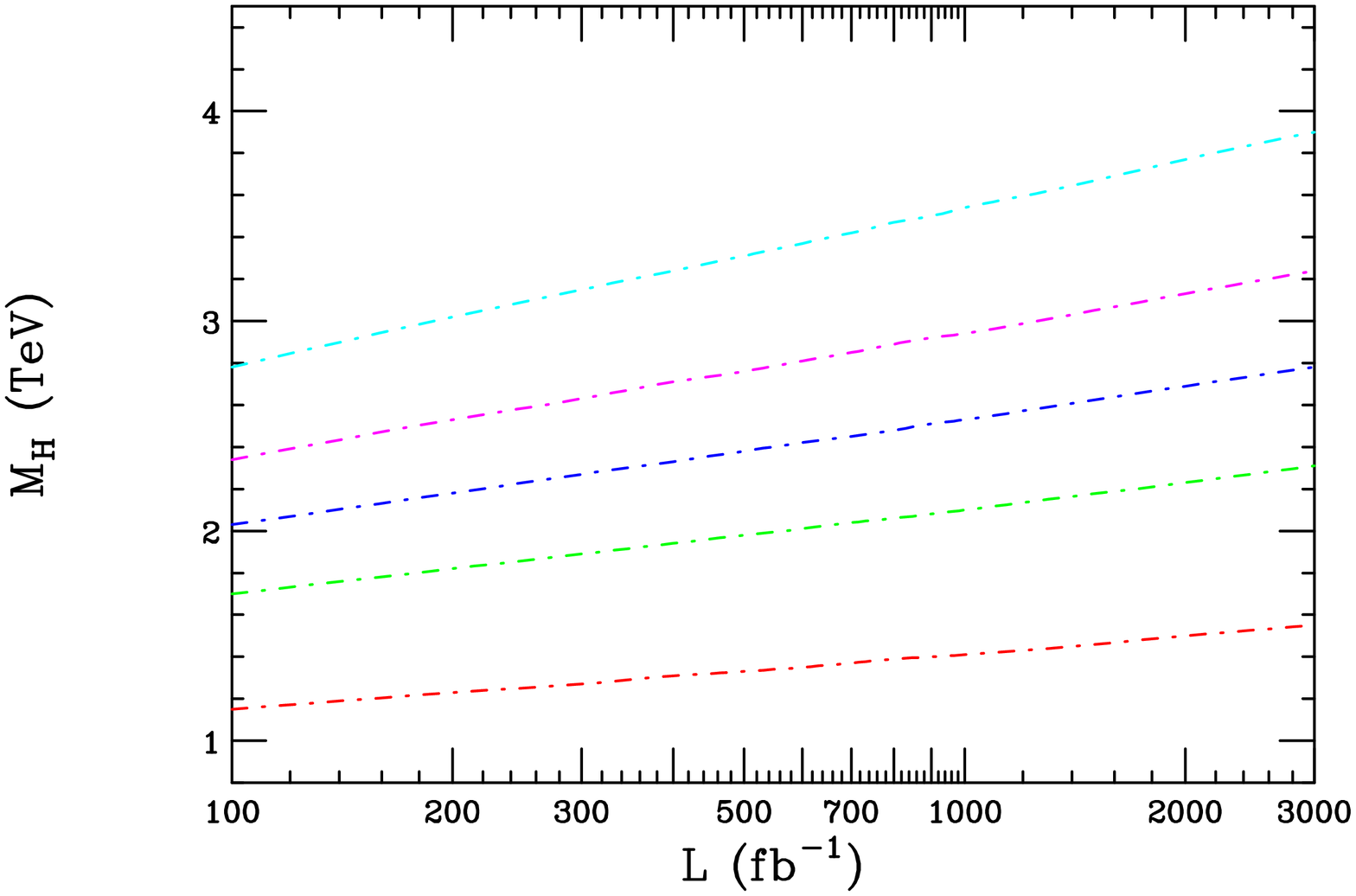}}
\vspace*{0.25cm}
\caption{$5\sigma$ reach for the discovery of a nonzero value of 
the azimuthal asymmetry 
$N^{-1} dA_i/dz$ distribution as a function of the integrated luminosity 
at a LC. The top(bottom) panel is for $\delta=3(4)$. Within each panel, 
from bottom to top the curves are for $\sqrt s=0.5, 0.8,1, 1.2$ and $1.5$ TeV, 
respectively. $M_H=M_D$ is assumed throughout as is $P_T=0.8$ and $P_T'=0.6$.}

\end{figure}

Assuming a value of $\delta$ we can ask up to what value of $M_H=M_D$ we can 
determine that the $N^{-1} dA_i/dz$ distribution is non-zero at the $5\sigma$ 
level. Based on the expression above we expect that this reach will be 
reasonably sensitive to the value of $\delta$; this is indeed what we find from 
Figs. 5-6 which show the resulting reaches at the $5\sigma$ level for the 
range $\delta=3-6$. The first thing to notice is that the sensitivity to this 
imaginary term decreases as $\delta$ is increased. This is not surprising 
since we see from the equation above that the magnitude of this term goes as 
$(\sqrt s/M_H)^{\delta-2}$ and $\sqrt s/M_H$ is always $<1$ while 
$\delta \geq 3$. For $\delta=3$ the reach is found to be $\sim (2.5-3)\sqrt s$ 
while for $\delta=6$ one obtains only $\sim 2\sqrt s$. Although these 
numbers are not large in comparison to those we've obtained in the 
other analyses above they 
provide the first indication that these two scenarios can be distinguished 
at a collider via indirect measurements. 

\vspace*{-0.5cm}
\nn
\begin{figure}[htbp]
\centerline{
\includegraphics[width=8cm,angle=90]{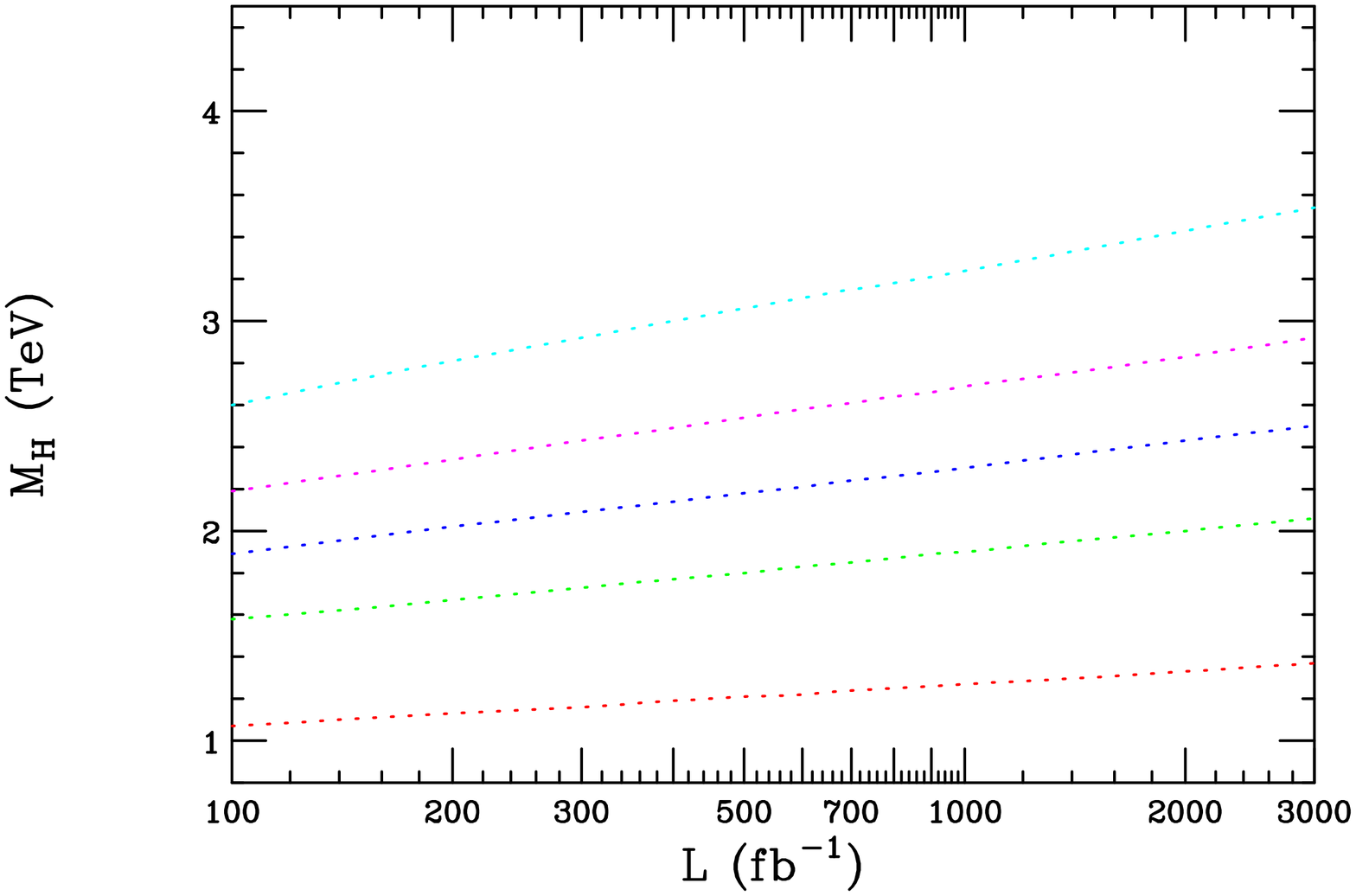}}
\vspace*{0.35cm}
\centerline{
\includegraphics[width=8cm,angle=90]{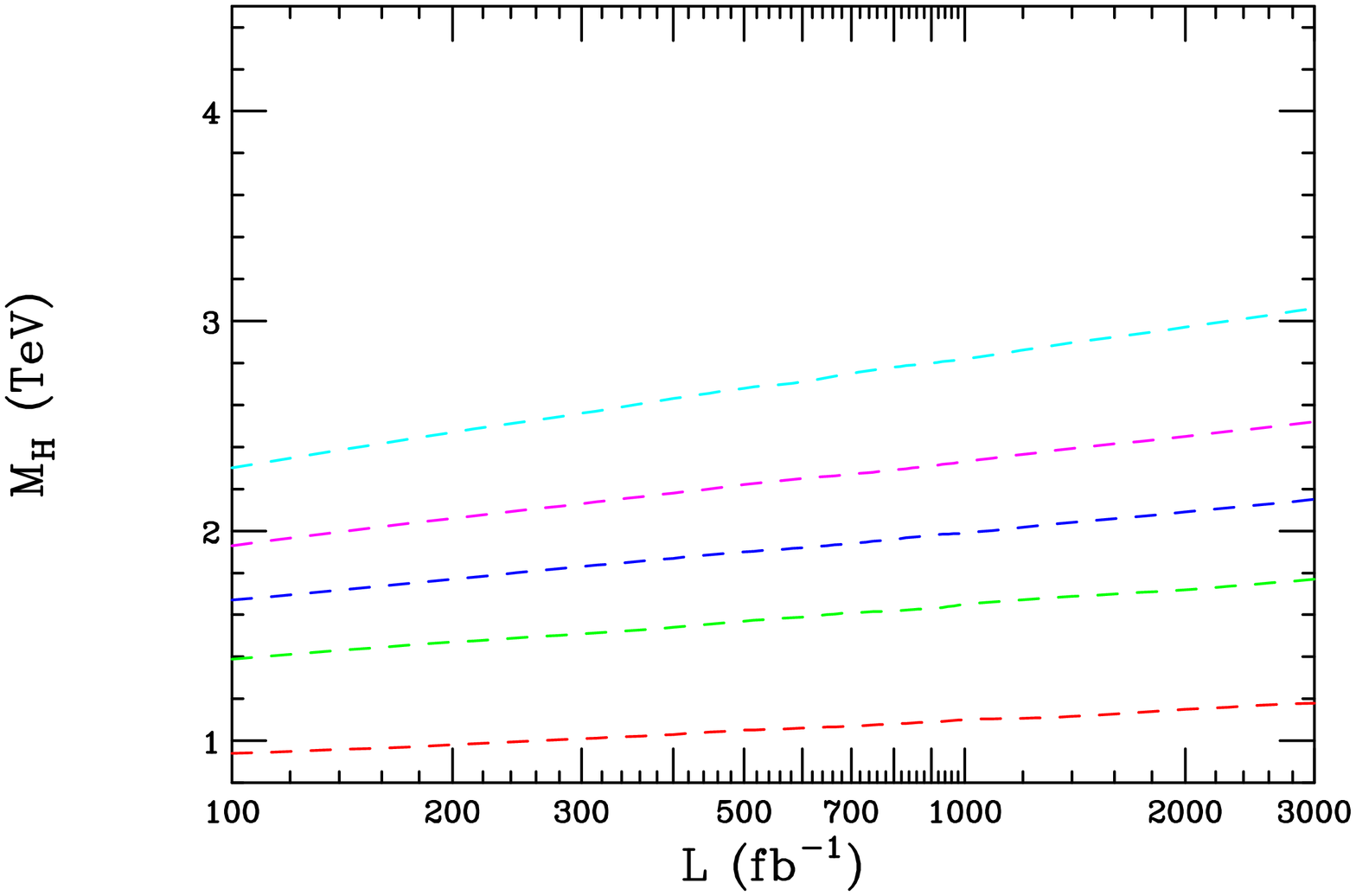}}
\vspace*{0.25cm}
\caption{Same as the previous figure but now for $\delta=5$(top) and 
$\delta=6$(bottom).}
\end{figure}

\section{Summary and Conclusion}

Historically, transverse polarization has not received much 
attention in the literature as a probe for new physics signatures. However, 
in searching for and identifying new physics at colliders one must make use 
of as many tools as possible.
 
In this paper we have examined the possible uses of transverse polarization 
in searching for, discovering and identifying spin-2 graviton exchange 
signatures in models with extra dimensions.  
The results of our analysis are as follows: ($i$) We have 
found that the interference of SM and spin-2 graviton KK exchanges leads to 
contributions to the azimuthal asymmetry distributions which are odd in 
$\cos \theta$, a rather unique signature. The appearance of such odd terms 
does not happen in the case of other new physics such as 
a $Z'$, contact interactions, gauge boson KK excitations or the exchange of 
new scalars. 
($ii$) Using two different sets of observables that probe the integrated 
contributions of these odd terms, we 
showed that it possible to differentiate KK graviton/spin-2 exchanges from 
all other new physics contributions to contact interactions at the $5\sigma$ 
level up to ADD cutoff scales of 
$M_H \sim (3.5-4)\sqrt s$. ($iii$) Fitting to the shape of the full 
differential distribution itself was shown to provide much more 
discriminating power; we found that the $5\sigma$ 
identification reach was substantially  increased to 
$M_H=(10-11)\sqrt s$, about a factor two improvement over what we obtained 
in our earlier analysis 
in the case of longitudinal polarization. This result is, however, quite 
sensitive to our assumptions about the sizes of various systematic errors. 
($iv$) Using this same
type of analysis we obtained a $95\%$ CL search reach for new physics in 
excess of $m_H=20\sqrt s$; this is again about a factor of two improvement over 
other analyses. As in the previous analysis, this result is also quite 
sensitive to the assumed values of the systematic errors. Clearly, more 
detailed studies are required to verify these results. 
($v$) In the case of the ADD model, an additional imaginary 
piece of the amplitude is present in comparison to the RS model below KK 
production threshold. We showed that this leads to a new asymmetry, produced 
through transverse polarization, which allows RS and ADD model 
separation at $5\sigma$ up to masses $M_H=(2.5-3)\sqrt s$. 
 
It is clear from our analysis that transverse polarization, though somewhat 
long neglected, can be a very powerful tool in identifying new physics, 
particularly in the case of extra dimensions. Further detailed study of the 
effects examined here may prove extremely useful for future linear collider 
experiments.

\noindent{\Large\bf Acknowledgements}

The author would like to thank J.L. Hewett for discussion related to this 
work. The author would also like to thank K. Desch for an inspirational talk 
on transverse polarization at the Jeju meeting that returned his interest 
to this subject after many years. 

%
\def\MPL #1 #2 #3 {Mod. Phys. Lett. {\bf#1},\ #2 (#3)}
\def\NPB #1 #2 #3 {Nucl. Phys. {\bf#1},\ #2 (#3)}
\def\PLB #1 #2 #3 {Phys. Lett. {\bf#1},\ #2 (#3)}
\def\PR #1 #2 #3 {Phys. Rep. {\bf#1},\ #2 (#3)}
\def\PRD #1 #2 #3 {Phys. Rev. {\bf#1},\ #2 (#3)}
\def\PRL #1 #2 #3 {Phys. Rev. Lett. {\bf#1},\ #2 (#3)}
\def\RMP #1 #2 #3 {Rev. Mod. Phys. {\bf#1},\ #2 (#3)}
\def\NIM #1 #2 #3 {Nuc. Inst. Meth. {\bf#1},\ #2 (#3)}
\def\ZPC #1 #2 #3 {Z. Phys. {\bf#1},\ #2 (#3)}
\def\EJPC #1 #2 #3 {E. Phys. J. {\bf#1},\ #2 (#3)}
\def\IJMP #1 #2 #3 {Int. J. Mod. Phys. {\bf#1},\ #2 (#3)}
\def\JHEP #1 #2 #3 {J. High En. Phys. {\bf#1},\ #2 (#3)}


\begin{thebibliography}{99}
%
\bibitem{e6}
J.~L.~Hewett and T.~G.~Rizzo,
Phys.\ Rept.\  {\bf 183}, 193 (1989).
%
\bibitem{zp}
A.~Leike, 
Phys.\ Rept.\  {\bf 317}, 143 (1999)
[arXiv:hep-ph/9805494];
M.~Cvetic and S.~Godfrey,
arXiv:hep-ph/9504216.
%
\bibitem{lq}
W.~Buchmuller, R.~Ruckl and D.~Wyler,
Phys.\ Lett.\ B {\bf 191}, 442 (1987)
[Erratum-ibid.\ B {\bf 448}, 320 (1999)]; 
J.~L.~Hewett and T.~G.~Rizzo,
Phys.\ Rev.\ D {\bf 58}, 055005 (1998)
[arXiv:hep-ph/9708419], 
Phys.\ Rev.\ D {\bf 56}, 5709 (1997)
[arXiv:hep-ph/9703337] and 
Phys.\ Rev.\ D {\bf 36}, 3367 (1987).
%
\bibitem{rp}
J.~Kalinowski, R.~Ruckl, H.~Spiesberger and P.~M.~Zerwas,
Phys.\ Lett.\ B {\bf 414}, 297 (1997)
[arXiv:hep-ph/9708272]; 
J.~Kalinowski, R.~Ruckl, H.~Spiesberger and P.~M.~Zerwas,
Phys.\ Lett.\ B {\bf 406}, 314 (1997)
[arXiv:hep-ph/9703436]; 
T.~G.~Rizzo,
Phys.\ Rev.\ D {\bf 59}, 113004 (1999)
[arXiv:hep-ph/9811440].
%
\bibitem{bl}
F.~Cuypers and S.~Davidson,
Eur.\ Phys.\ J.\ C {\bf 2}, 503 (1998)
[arXiv:hep-ph/9609487].
%
\bibitem{ed}
J.~L.~Hewett,
Phys.\ Rev.\ Lett.\  {\bf 82}, 4765 (1999)
[arXiv:hep-ph/9811356]; 
G.~F.~Giudice, R.~Rattazzi and J.~D.~Wells,
Nucl.\ Phys.\ B {\bf 544}, 3 (1999)
[arXiv:hep-ph/9811291];
T.~Han, J.~D.~Lykken and R.~J.~Zhang,
Phys.\ Rev.\ D {\bf 59}, 105006 (1999)
[arXiv:hep-ph/9811350];
T.~G.~Rizzo,
Phys.\ Rev.\ D {\bf 59}, 115010 (1999)
[arXiv:hep-ph/9901209].
%
\bibitem{dhr} 
For an overview of RS phenomenology, see 
H.~Davoudiasl, J.~L.~Hewett and T.~G.~Rizzo, 
Phys.\ Rev.\ D {\bf 63}, 075004 (2001)
[arXiv:hep-ph/0006041]; Phys.\ Lett.\ B {\bf 473}, 43 (2000)
[arXiv:hep-ph/9911262]; Phys.\ Rev.\ Lett.\  {\bf 84}, 2080 (2000)
[arXiv:hep-ph/9909255];
J.~L.~Hewett, F.~J.~Petriello and T.~G.~Rizzo, 
JHEP {\bf 0209}, 030 (2002) [arXiv:hep-ph/0203091].
%
\bibitem{add}
N.~Arkani-Hamed, S.~Dimopoulos and G.~R.~Dvali,
Phys.\ Lett.\ B {\bf 429}, 263 (1998)
[arXiv:hep-ph/9803315];
I.~Antoniadis, N.~Arkani-Hamed, S.~Dimopoulos and G.~R.~Dvali,
Phys.\ Lett.\ B {\bf 436}, 257 (1998)
[arXiv:hep-ph/9804398];
N.~Arkani-Hamed, S.~Dimopoulos and G.~R.~Dvali,
Phys.\ Rev.\ D {\bf 59}, 086004 (1999)
[arXiv:hep-ph/9807344].
%
\bibitem{rs} 
L. Randall and R. Sundrum, \PRL 83 3370 1999 .  
%
\bibitem{ed2}
See, for example, 
I.~Antoniadis,
Phys.\ Lett.\ B {\bf 246}, 377 (1990); 
T.~G.~Rizzo and J.~D.~Wells,
Phys.\ Rev.\ D {\bf 61}, 016007 (2000)
[arXiv:hep-ph/9906234];
M.~Masip and A.~Pomarol,
Phys.\ Rev.\ D {\bf 60}, 096005 (1999)
[arXiv:hep-ph/9902467];
T.~G.~Rizzo,
Phys.\ Rev.\ D {\bf 64}, 015003 (2001)
[arXiv:hep-ph/0101278].
%
\bibitem{se}
S.~Cullen, M.~Perelstein and M.~E.~Peskin,
Phys.\ Rev.\ D {\bf 62}, 055012 (2000)
[arXiv:hep-ph/0001166].
%
\bibitem{gaby}
G.~Pasztor and M.~Perelstein,
in {\it Proc. of the APS/DPF/DPB Summer Study on the Future of Particle 
Physics (Snowmass 2001) } ed. R.~Davidson and C.~Quigg,
arXiv:hep-ph/0111471.

%
\bibitem{tp}
R. Budny, \PRD D14 2969 1976 ; 
H.A. Olsen, P. Osland and I. Overbo, \PLB B97 286 1980 ;
K. Hikasa, \PRD D33 3203 1986 ; 
C.P. Burgess and J.A. Robinson, \IJMP A6 2709 1991 ;
A. Djouadi, F.M. Renard and C. Verzegnassi, \PLB B241 260 1990 ;
F.M. Renard, \ZPC C44 75 1989 ;
J.L. Hewett and T.G. Rizzo, \ZPC C44 75 1987 ~and \ZPC C36 209 1987 ;
J. Fleischer, K. Kolodziej and F. Jegerlehner, \PRD D49 2174 1994 ;
for a recent discussion of this option at the LC, see K. Desch, talk given 
at the {\it International Workshop on the Linear Collider, LCWS2002}, Jeju 
Island, Korea, Aug. 2002.  
%
\bibitem{gudi}
G. Moortgat-Pick, private communication.
%
\bibitem{tgr}
T.G. Rizzo, JHEP 0210 013 2002.
%
\bibitem{mp}
E. Eichten, K. Lane and M. Peskin, \PRL 50 811 1983 .
%
\bibitem{tesla}
J.~A.~Aguilar-Saavedra {\it et al.}  [ECFA/DESY LC Physics Working Group
                  Collaboration],
``TESLA Technical Design Report Part III: Physics at an e+e- Linear Collider,''
arXiv:hep-ph/0106315; 
S. Riemann, TESLA Linear Collider Note LC-TH-2001-007'
%
\bibitem{joa}
For a review of the current and anticipated future collider reaches for 
extra dimensions, see
J.~Hewett and M.~Spiropulu,
``Particle physics probes of extra spacetime dimensions,''
arXiv:hep-ph/0205106.
%
\bibitem{datta}
A. Datta, E. Gabrielli and B. Mele, hep-ph/0210318. 

\end{thebibliography}
\end{document}